\documentclass[a4paper, 11pt]{article}
\usepackage[T1]{fontenc}
\usepackage{jheppub,graphicx,epsf,amssymb,amsbsy,amsfonts,amssymb,amsmath, empheq}
\def\be{\begin{equation}}
\def\ee{\end{equation}}
\def\bea{\begin{eqnarray}}
\def\eea{\end{eqnarray}}

\def\bg{\bar{g}}
\def\beq{\begin{eqnarray}}\def\eeq{\end{eqnarray}}
\def\ba#1\ea{\begin{align}#1\end{align}}
\def\bg#1\eg{\begin{gather}#1\end{gather}}
\def\bm#1\em{\begin{multline}#1\end{multline}}
\def\bmd#1\emd{\begin{multlined}#1\end{multlined}}

\def\D{\Delta}

\def\({\left(}
\def\){\right)}
\def\[{\left[}
\def\]{\right]}

\def\D{\Delta}

\title{(Chiral) Virasoro invariance of the tree-level MHV graviton scattering amplitudes}

\author[1]{Shamik Banerjee}
\author[2]{Sudip Ghosh}
\author[3]{Partha Paul}
\affiliation[1]{Institute of Physics, Sachivalaya Marg, Bhubaneshwar, India-751005 \\ and Homi Bhabha National Institute, Anushakti Nagar, Mumbai, India-400085}
\affiliation[2]{Okinawa Institute of Science and Technology,1919-1 Tancha, Onna-son, Okinawa 904-0495,Japan}
\affiliation[3]{Chennai Mathematical Institute, SIPCOT IT Park, Siruseri 603103, India}
\emailAdd{banerjeeshamik.phy@gmail.com, sudip112phys@gmail.com, pl.partha13@gmail.com }

\abstract{In this paper we continue our study of the tree level MHV graviton scattering amplitudes from the point of view of celestial holography. In arXiv:2008.04330 we showed that the celestial OPE of two gravitons in the MHV sector can be written as a linear combination of $\overline{SL(2,\mathbb C)}$ current algebra and supertranslation descendants. In this note we show that the OPE is in fact manifestly invariant under the infinite dimensional Virasoro algebra as is expected for a $2$-D CFT. This is consistent with the conjecture that the holographic dual in $4$-D asymptotically flat space time is a $2$-D CFT. Since we get only one copy of the Virasoro algebra we can conclude that the holographic dual theory which computes the MHV amplitudes is a chiral CFT with a host of other infinite dimensional global symmetries including $\overline{SL(2,\mathbb C)}$ current algebra, supertranslations and subsubleading soft graviton symmetry. We also discuss some puzzles related to the appearance of the Virasoro symmetry.}

\begin{document}
\maketitle
\flushbottom

\section{Introduction and Results}
In four dimensional asymptotically flat space time the asymptotic symmetries are infinite dimensional \cite{Sachs:1962zza,Strominger:2013jfa,He,Barnich:2009se,Barnich:2011ct,Kapec:2016jld,Kapec:2014opa,He:2017fsb,Banerjee:2020zlg,Guevara:2021abz,Banerjee:2021cly,Strominger:2021lvk,Donnay:2020guq,Bagchi:2016bcd} and so in a holographic description the dual theory must have infinite dimensional global symmetry. It has been conjectured \cite{Barnich:2009se,Barnich:2011ct,Kapec:2016jld,Kapec:2014opa,He:2017fsb,Bagchi:2016bcd} that the global symmetry group of the dual theory contains the local conformal or Virasoro symmetry. As a result the dual theory is a two dimensional CFT. In this note we study this conjecture in the context of celestial holography \cite{Pasterski:2017kqt}. 

We revisit the global symmetries of the tree level MHV graviton scattering amplitudes calculated in General Relativity. In \cite{Banerjee:2020zlg}  we studied the celestial MHV amplitudes in great detail and showed that the celestial OPE between two graviton primaries in the MHV sector can be written as a linear combination of $\overline{SL(2,\mathbb C)}$ current algebra, supetranslations and global $SL(2,\mathbb C)$ descendants. Moreover, due to the existence of the null states \cite{Banerjee:2020zlg}\footnote{The generators of the $\overline{SL(2,\mathbb C)}$ current algebra are denoted by $J^a_n$ where $a = 0,\pm 1$ and $n\in \mathbb Z$. Similarly we denote the generators of the supertranslations coming from the positive helicity soft graviton by $P_{n,b}$ where $n\in \mathbb Z$ and $b = 0,-1$. For further details on the symmetry algebra and null states we refer the reader to \cite{Banerjee:2020zlg}.},
\be
\Psi_{\D}(z,\bar z) = \[J^1_{-1} P_{-1,-1} - (\Delta-1) P_{-2,0}\]G^{+}_{\D}(z,\bar{z}) = 0 
\ee
and
\be
\Phi_{\Delta}(z,\bar z) = \left[L_{-1} P_{-1,-1}+ 2 J^{0}_{-1} P_{-1,-1}- (\Delta+1) P_{-2,-1} - \bar{L}_{-1}P_{-2,0} \right] G^{+}_{\Delta}(z,\bar{z}) =0
\ee
one can write down differential equations which can be solved to determine the MHV  graviton scattering amplitudes. Therefore the MHV amplitudes are completely determined by the $\overline{SL(2,\mathbb C)}$ current algebra, supetranslations and the global $SL(2,\mathbb C)$ symmetries. 

In this paper we point out that the celestial OPE of two graviotns in the MHV sector is actually \textit{invariant under the infinite dimensional \textbf{Viarasoro} algebra} whose global part is the $SL(2,\mathbb C)$ algebra. This shows that the symmetries of the MHV graviton scattering amplitudes is a \textit{semi-direct product of the Virasoro algebra, $\overline{SL(2,\mathbb C)}$ current algebra and supertranslations}.\footnote{See \cite{paul} for a potential derivation of this symmetry algebra from the bulk gravity point of view.} Since we get only one copy of the Virasoro algebra, the \textbf{holographic dual theory} which computes the MHV graviton scattering amplitudes is a \textbf{chiral (celestial) CFT}.

For the sake of completeness let us now write down the symmetry algebra,

\subsection{Virasoro algebra}
\be
\[ L_m, L_n\] = (m-n) L_{m+n} + \textcolor{red}{\frac{c}{12} n(n^2-1) \delta_{m+n,0}}   \quad m,n \in \mathbb Z
\ee

Although we have written down the central charge term in the Virasoro algebra, our demonstration of Virasoro invariance of the celestial OPE does not allow us to determine the value of the central charge $c$. So we leave the determination of $c$ to future work. 


\subsection{$\overline{SL(2,\mathbb C)}$ current algebra}
\be
\[ J^a_m, J^b_n\] = (a-b) J^{a+b}_{m+n}, \quad a,b = 0, \pm 1, \quad m,n \in \mathbb{Z}
\ee

\be
J^1_0 = \bar L_1, \quad J^0_0 = \bar L_0, \quad J^{-1}_0 = \bar L_{-1}
\ee


\subsection{Supertranslation}
\be
\[ P_{m,n}, P_{m',n'}\] = 0, \quad n, n' = 0, -1
\ee

\subsection{Mixed commutators}
\be
\quad [J^1_m, P_{n,-1}] =  P_{m+n,0}, \quad  [J^0_m, P_{n,-1}] = \frac{1}{2} P_{m+n,-1}, \quad [J^{-1}_m, P_{n,-1}] = 0 
\ee 

\be
\quad \[ J^1_m, P_{n,0}\] =0, \quad \[J^0_m, P_{n,0}\] = - \frac{1}{2}P_{m+n,0} , \quad [J^{-1}_m, P_{n,0}] = - P_{m+n,-1} 
\ee

\be
[L_m, J^a_n] = -n J^a_{m+n}, \quad m \in \mathbb Z, \quad n\in \mathbb{Z}
\ee

\be
\[ L_n, P_{a,b}\] = \( \frac{n-1}{2} - a\) P_{a+n,b}, \quad n \in \mathbb Z, \quad a \in \mathbb Z,  \quad b =0, -1
\ee

The symmetry algebra that we have shown can be further extended by including the subsubleading soft graviton symmetry \cite{Guevara:2021abz,Banerjee:2021cly,Strominger:2021lvk} and an infinite number of other soft symmetries \cite{Guevara:2021abz,Strominger:2021lvk} which appear when the scaling dimension of a (positive helicity) graviton primary assumes negative integer values.

\section{Virasoro Invariance of the OPE in the MHV sector}

Consider the celestial OPE between two positive helicity outgoing gravitons in the MHV sector. This is given by \cite{Banerjee:2020zlg}, 
\begin{equation}
\label{ppope}
\begin{split}
  & G^{+}_{\Delta_{1}} (z,\bar{z}) G^{+}_{\Delta_{2}} (0) \Big|_{MHV}  \\
  &=   \frac{\bar{z}}{z}  \ B(\Delta_{1}-1,\Delta_{2}-1) \bigg[ - P_{-1,-1}    +  z \big(c_{1} \hspace{0.04cm} J^{0}_{-1}P_{-1,-1} + c_{2} P_{-2,-1}  \big) \\
 &+ z^{2} \bigg(  c_{3}  \hspace{0.04cm} J^{0}_{-2} P_{-1,-1} + c_{4}  P_{-3,-1}  + c_{5} \big( 2 L_{-1}P_{-2,-1} - 2 \bar{L}_{-1}P_{-3,0}+ 2L_{-1}\bar{L}_{-1} P_{-2,0} -  L^{2}_{-1} P_{-1,-1} \big) \bigg) \\
 & +\mathcal O(z^3) \bigg] G^{+}_{\Delta_{1}+\Delta_{2}-1} (0)   + \cdots
\end{split}
\end{equation}

The descendant OPE coefficients denoted above by $c_{i}$ are given by
\begin{equation}
\label{ppopeci}
\begin{split}
  & c_{1} = \frac{2(\Delta_{1}-1)}{\Delta_{1}+\Delta_{2}-2} , \quad c_{2} = - \Delta_{1}, \quad c_{3} = \frac{2(\Delta_{1}-1) (\Delta_{2}-1)}{(\Delta_{1}+\Delta_{2}-2)(\Delta_{1}+\Delta_{2}-1)} , \\
  & c_{4} = - \frac{\Delta_{1}\Delta_{2}}{\Delta_{1}+\Delta_{2}-1}, \quad c_{5} = \frac{\Delta_{1}(\Delta_{1}-1) }{2(\Delta_{1}+\Delta_{2}-2)(\Delta_{1}+\Delta_{2}-1)} 
\end{split}
\end{equation}

In \eqref{ppope} we have, for simplicity, kept only terms of the $\mathcal O(\bar z z^n)$ where $n\ge -1$ because under the action of the Virasoro generators these terms do not mix with other terms in the OPE. 

We already know \cite{Banerjee:2020zlg} that the OPE \eqref{ppope} is invariant under the action of the global generator $L_1$. In this section we show that both sides of the OPE \eqref{ppope} transform in the same way under the action of the first Virasoro generator $L_2$. This check is non-trivial because the OPE \eqref{ppope} has terms of $\mathcal O(\bar z z)$. 

It is important to note that the $\mathcal O(\bar z z)$ term in \eqref{ppope} does not contain the $L_{-2}$ descendant of the graviton primary and so the Virasoro central charge $c$ does not play any role in the process of checking the Virasoro invariance. Let us now spell out the details. 

\subsection{$L_2$ invariance} 
Let us start with the OPE 
\begin{equation}
\label{ppope1}
\begin{split}
  & G^{+}_{\Delta_{1}} (z,\bar{z}) G^{+}_{\Delta_{2}} (0) \Big|_{MHV}  \\
  &=   \frac{\bar{z}}{z}  \ B(\Delta_{1}-1,\Delta_{2}-1) \bigg[ - P_{-1,-1}    +  z \big(c_{1} \hspace{0.04cm} J^{0}_{-1}P_{-1,-1} + c_{2} P_{-2,-1}  \big) \\
 &+ z^{2} \bigg(  c_{3}  \hspace{0.04cm} J^{0}_{-2} P_{-1,-1} + c_{4}  P_{-3,-1}  + c_{5} \big( 2 L_{-1}P_{-2,-1} - 2 \bar{L}_{-1}P_{-3,0}+ 2L_{-1}\bar{L}_{-1} P_{-2,0} -  L^{2}_{-1} P_{-1,-1} \big) \bigg) \\
 & +\mathcal O(z^3) \bigg] G^{+}_{\Delta_{1}+\Delta_{2}-1} (0)   + \cdots
\end{split}
\end{equation}

We now show that both sides of \eqref{ppope1} transform in the same way under the action of the Virasoro generator $L_{2}$. For this we take the gravitons in the OPE to be Virasoro primaries so that they satisfy \footnote{Here $(h,\bar h)$ denote the holomorphic and antiholomorphic conformal dimensions given by $h= \frac{\D+\sigma}{2}$ and $\bar h = \frac{\D-\sigma}{2}$ where $\sigma$ is the helicity of the graviton.}
\begin{equation}
\label{L2def}
\begin{split}
  & [ L_{2}, G_{h,\bar{h}} (z,\bar{z}) ] = z^{2} (z\partial_{z} + 3 h) G_{h,\bar{h}} (z,\bar{z})
\end{split}
\end{equation}

Now we apply $L_2$ to the LHS of \eqref{ppope1} and obtain using \eqref{L2def}
\begin{equation}
\label{L2actpplhs}
\begin{split}
  & [ L_{2}, G^{+}_{\Delta_{1}} (z,\bar{z}) G^{+}_{\Delta_{2}} (0)  ] = z^{2} (z\partial_{z} + 3 h_{1}) G^{+}_{\Delta_{1}} (z,\bar{z}) G^{+}_{\Delta_{2}} (0) \sim \mathcal O(\bar z z) + \cdots
\end{split}
\end{equation}

where the dots denote terms of $\mathcal O(\bar z z^2)$ and higher. The leading $\mathcal O(\bar z z)$ term in \eqref{L2actpplhs} is given by,
\begin{equation}
\label{ppL2zzbar1}
\begin{split}
   [ L_{2}, G^{+}_{\Delta_{1}} (z,\bar{z}) G^{+}_{\Delta_{2}} (0)  ] \Big|_{\mathcal{O}(z\bar{z})}  & = z\bar{z} \ B(\Delta_{1}-1,\Delta_{2}-1) \ (1-3h_{1}) P_{-1,-1} G^{+}_{\Delta_{1}+\Delta_{2}-1} (0)  \\
   & = - z\bar{z} \ B(\Delta_{1}-1,\Delta_{2}-1) \ \frac{(3\Delta_{1}+4)}{2} P_{-1,-1} G^{+}_{\Delta_{1}+\Delta_{2}-1} (0) 
\end{split}
\end{equation}

Now we apply to $L_2$ to the RHS of the OPE \eqref{ppope1}. Using the following commutation relations
\begin{equation}
\label{comms}
\begin{split}
  & [ L_{m}, J^{a}_{n}] = -n J^{a}_{m+n}, \quad [L_{m}, P_{r,s}] = \left( \frac{m-1}{2} -r \right) P_{r+m,s}, \\
  & [J^{a}_{m}, J^{b}_{n}] = (a-b)J^{a+b}_{m+n}, \quad [P_{m,n}, P_{r,s}] =0  
\end{split}
\end{equation}

and the values of the coefficients $c_{3},c_{4},c_{5}$ given in \eqref{ppopeci} we find that the action of $L_{2}$ on the R.H.S. of \eqref{ppope1} yields
\begin{equation}
\label{ppL2zzbar2}
\begin{split} 
& z \bar{z} \ B(\Delta_{1}-1,\Delta_{2}-1) \left[ (\Delta_{1}+\Delta_{2}-2)c_{3} + \frac{7 c_{4}}{2} - 3c_{5}(\Delta_{1}+ \Delta_{2}-2)\right] P_{-1,-1} G^{+}_{\Delta_{1}+\Delta_{2}-1} (0) + \cdots \\
  & = - z \bar{z} \ B(\Delta_{1}-1,\Delta_{2}-1) \ \frac{(3\Delta_{1}+4)}{2} P_{-1,-1} G^{+}_{\Delta_{1}+\Delta_{2}-1} (0) + \cdots
\end{split}
\end{equation}

where the dots denote terms higher order in $z$. We can see that the $\mathcal O(\bar z z)$ terms in \eqref{ppL2zzbar1} and \eqref{ppL2zzbar2} match precisely as expected from the Virasoro invariance of the OPE \eqref{ppope}. 

In the appendix we perform more checks of the Virasoro invariance of the celestial OPE of $++$ and $+-$ helicity gravitons in the MHV sector.  



\subsection{Reorganising the OPE in terms of Virasoro primaries and descendants}
\label{virrep1}

We have shown that up to $\mathcal O(\bar z z)$ terms the OPE of positive helicity outgoing gravitons in the MHV sector given by \eqref{ppope} is invariant under the (holomorphic) Virasoro symmetry. This implies that this OPE can be reorganised according to representations of the Virasoro symmetry algebra as is usually done in $2$-$d$ CFTs. In other words, we want to show that every term at a particular order in the OPE is either a Virasoro primary or a Virasoro descendant of another Virasoro primary which appears at a lower order in $z$. The OPE coefficient multiplying a Virasoro primary cannot be obtained by using the Virasoro symmetry alone but the OPE coefficient multiplying a Virasoro descendant is uniquely determined by the Virasoro symmetry and the OPE coefficient of the Virasoro primary from which it descends. For simplicity here we reorganise only the $\mathcal O(\bar z z^n)$ terms in the OPE.

\vskip 4pt
Now in terms of Virasoro primaries and descendants, we find that the positive helicity OPE \eqref{ppope} can be expressed as
\begin{equation}
\label{ppvirrep}
\begin{split}
  & G^{+}_{\Delta_{1}} (z,\bar{z}) G^{+}_{\Delta_{2}} (0) \Big|_{MHV} \\
  & = \frac{\bar{z}}{z} \  B(\Delta_{1}-1,\Delta_{2}-1) \bigg[ C_{12\phi_{1}}  \left( 1 + z \hspace{0.04cm} \beta_{\phi_{1}}^{(1)} L_{-1}+ z^{2}\beta_{\phi_{1}}^{(1,1)} L_{-1}^{2}+ z^{2}\beta_{\phi_{1}}^{(2)}\textcolor{blue}{L_{-2}} \right) \phi_{1}(0)  \\
   & +  z \hspace{0.03cm}  C_{12\phi_{2}}  \left( 1 + z \hspace{0.04cm}  \beta_{\phi_{2}}^{(1)} L_{-1} \right) \phi_{2}(0)   + z \hspace{0.03cm} C_{12\phi_{3}}  \left( 1 + z \hspace{0.04cm}  \beta_{\phi_{3}}^{(1)} L_{-1} \right) \phi_{3}(0)  + z^{2} \hspace{0.03cm} C_{12\phi_{4}}   \phi_{4}(0) + \cdots \bigg] + \cdots
\end{split}
\end{equation}

where $\phi_{i}(0)$ with $i=1,2,3,4$ are Virasoro primaries. Thus they satisfy the conditions
\begin{equation}
\label{virprims}
L_0 \phi_i(0) = h_{\phi_i} \phi_i(0), \quad L_{m} \phi_{i}(0) = 0,  \quad m \ge 1 
\end{equation}
where $h_{\phi_i}$ is the scaling dimension of the Virasoro primary $\phi_i(0)$.

The coefficients $C_{12\phi_i}$ in \eqref{ppvirrep} are given by
 \begin{equation}
\label{C12p} 
\begin{split}
& C_{12\phi_{1}}  = - 1, \quad C_{12\phi_{2}} = \frac{\Delta_{1}-1}{\Delta_{1}+\Delta_{2}-2}, \quad C_{12\phi_{3}}= \frac{\Delta_{1}-\Delta_{2}}{\Delta_{1}+\Delta_{2}-2}, \quad C_{12\phi_{4}} = -  \frac{\Delta_{1}\Delta_{2} }{\Delta_{1}+\Delta_{2}-1}
\end{split}
\end{equation}

The $\beta_{\phi_i}^{(\vec k)}$'s  in \eqref{ppvirrep} are completely determined by the Virasoro algebra and are given by  
\begin{equation}
\label{betas}
\begin{split}
  &  \beta_{\phi_i}^{(1)} = \frac{h_{1}-h_{2}+ h_{\phi_i}}{2 h_{\phi_i}}, \\
  & \beta_{\phi_i}^{(1,1)} =  \frac{4h_{\phi_i}\left(2 (h_{1}-h_{2}+h_{\phi_i})^{2} +h_{2}-4h_{1}-h_{\phi_i}\right)+c(h_{1}-h_{2}+h_{\phi_i})(h_{1}-h_{2}+h_{\phi_i}+1)}{4h_{\phi_i}\left(c+2 c h_{\phi_i} + 2 h_{\phi_i} (8 h_{\phi_i}-5)\right)} \\
  & \beta_{\phi_i}^{(2)}  = \frac{h_{\phi_i}(2h_{1}+2h_{2}+h_{\phi_i}-1)+ h_{1}+h_{2} -3(h_{1}-h_{2})^{2}}{\left( c+ 2 c h_{\phi_i} + 2 h_{\phi_i} (8 h_{\phi_i}-5)\right)}
  \end{split}
\end{equation}

where $h_{1},h_{2}$ are the holomorphic weights of the primary operators whose OPE is being considered and $h_{\phi_i}$ is the holomorphic weight of the Virasoro primary $\phi_{i}(0)$ which appears in this OPE. In our case 
\begin{equation}
\label{h1h2def}
\begin{split}
  &  h_{1} = \frac{\Delta_{1}+2}{2}, \quad h_{2} = \frac{\Delta_{2}+2}{2}
  \end{split}
\end{equation}


\vskip 4pt
Finally, the explicit expressions for the Virasoro primaries $\phi_i(0)$ are as follows
\begin{equation}
\label{phipsi}
\begin{split}
  &   \phi_{1}(0) = P_{-1,-1}G^{+}_{\Delta_{1}+\Delta_{2}-1} (0) , \quad   \phi_{2}(0) = \bar{L}_{-1}P_{-2,0}G^{+}_{\Delta_{1}+\Delta_{2}-1} (0),  \\
  &  \phi_{3}(0) = \left(P_{-2,-1}- \frac{2}{(\Delta_{1}+\Delta_{2}+2)} L_{-1}P_{-1,-1}\right)G^{+}_{\Delta_{1}+\Delta_{2}-1} (0)\\
  & \phi_{4}(0) = \bigg(P_{-3,-1} + a_{1} J^{0}_{-2}P_{-1,-1}+ a_{2}\textcolor{blue}{L_{-2}}P_{-1,-1}+ a_{3}L^{2}_{-1}P_{-1,-1} \\
  & \hspace{2.0cm}+ a_{4}L_{-1}P_{-2,-1}+ a_{5} L_{-1}\bar{L}_{-1}P_{-2,0}+ a_{6}\bar{L}_{-1}P_{-3,0}\bigg)G^{+}_{\Delta_{1}+\Delta_{2}-1} (0)
    \end{split}
\end{equation}

where the coefficients denoted by $a_{i}$ and $b_{i}$ are given by
\begin{equation}
\label{ais}
\begin{split}
  &  a_{1} = \frac{2(\Delta_{1}-1) (\Delta_{2}-1)}{C_{12\phi_{4}}(\Delta_{1}+\Delta_{2}-2)(\Delta_{1}+\Delta_{2}-1)}, \quad a_{2} = -\frac{C_{12\phi_{1}} \ \beta_{\phi_{1}}^{(2)}}{C_{12\phi_{4}}},   \\
  & a_{3} = \frac{1}{C_{12\phi_{4}}} \left( - \frac{\Delta_{1}(\Delta_{1}-1) }{2(\Delta_{1}+\Delta_{2}-2)(\Delta_{1}+\Delta_{2}-1)} - C_{12\phi_{1}} \ \beta_{\phi_{1}}^{(1,1)}  +   \frac{2 \hspace{0.03cm} C_{12\phi_{3}} \ \beta_{\phi_{3}}^{(1)}}{(\Delta_{1}+\Delta_{2}+2)}  \right),  \\
  &  a_{4} =  \frac{1}{C_{12\phi_{4}}} \left(  \frac{\Delta_{1}(\Delta_{1}-1) }{(\Delta_{1}+\Delta_{2}-2)(\Delta_{1}+\Delta_{2}-1)} - C_{12\phi_{3}} \ \beta_{\phi_{3}}^{(1)} \right),  \\
  & a_{5} =  \frac{1}{C_{12\phi_{4}}} \left(  \frac{\Delta_{1}(\Delta_{1}-1) }{(\Delta_{1}+\Delta_{2}-2)(\Delta_{1}+\Delta_{2}-1)} - C_{12\phi_{2}} \ \beta_{\phi_{2}}^{(1)} \right),  \\
  & a_{6} = -   \frac{\Delta_{1}(\Delta_{1}-1) }{C_{12\phi_{4}}(\Delta_{1}+\Delta_{2}-2)(\Delta_{1}+\Delta_{2}-1)}  
   \end{split}
\end{equation}

As we have already mentioned we are not able to determine the value of the Virasoro central charge $c$ in this way. The reason being that the $\mathcal O(\bar z z)$ term in the OPE \eqref{ppope} does not contain the $L_{-2}$ descendant of the graviton primary and so both sides of the OPE \eqref{ppope} transform in the same way under $L_2$ irrespective of the value of the central charge $c$. This is also evident from the decomposition \eqref{ppvirrep} which holds for \textit{arbitrary}
value of the central charge $c$. In particular the states $\phi_i(0)$ are Virasoro primaries for any value of the central charge as can be easily checked. We leave the determination of the value of the central charge to future research. \footnote{Since we are considering tree level scattering amplitudes, presumably the value of the central charge $c$ is zero.}

\section{Discussion}
Since the MHV sector is manifestly Virasoro invariant, the dual theory which computes the MHV scattering amplitudes has a stress tensor $T(z)$ whose modes are the Virasoro generators $L_n$. In celestial CFT the stress tensor is given by \cite{Kapec:2014opa,Kapec:2016jld,He:2017fsb} the shadow of the subleading conformally soft graviton. In particular the holomorphic stress tensor $T(z)$ is the shadow of the \textit{negative helicity} subleading soft graviton. Let us try to explain the appearance of $T(z)$ in the MHV sector along this line. 

So consider the NMHV amplitudes of the form $\langle{- - - ++\cdots +}\rangle$. Now if we make one of the negative helicity gravitons subleading (conformally) soft \cite{Cachazo:2014fwa, Donnay:2018neh} then the NMHV amplitude soft factorises via the MHV amplitude $\langle{- - ++\cdots +}\rangle$ and this can explain the appearance of the stress tensor $T(z)$ in the MHV sector. But for various reasons this explanation cannot be complete. For example, the negative helicity subleading soft graviton gives rise to $SL(2,\mathbb C)$ current algebra and its shadow gives the holomorphic stress tensor $T(z)$. Although we find the stress tensor $T(z)$ in the MHV sector, the $SL(2,\mathbb C)$ current algebra is not visible in the MHV sector. It is not clear that why these two symmetry algebras arising from the negative helicity subleading soft graviton are not treated on an equal footing. Similarly, we also have positive helicity soft graviton in the MHV sector and we do find the $\overline{SL(2,\mathbb C)}$ current algebra symmetry \cite{Banerjee:2020zlg} in the MHV sector but not the $\overline{\text{Virasoro}}$ symmetry. It may be that not all symmetries are manifest \cite{Strominger:2021lvk} at the level of OPE. We hope to return to these issues in future.    

\section{Acknowledgements}
We are grateful to Andy Strominger for very helpful correspondence on related topics. SB would also like to thank the string theory group in IISER Bhopal, India for their hospitality during the final stages of the work. SG would like to thank Yasha Neiman for useful discussions. The work of SB is partially supported by the Science and Engineering Research Board (SERB) grant MTR/2019/000937 (Soft-Theorems, S-matrix and Flat-Space Holography). The work of SG is supported by the Quantum Gravity Unit of the Okinawa Institute of Science and Technology Graduate University (OIST). The work of PP is partially supported by a grant to CMI from the Infosys Foundation.

\appendix

\section{Virasoro invariance of positive helicity OPE: further checks}
In this section, we provide additional checks of the Virasoro invariance of the positive helicity graviton OPE in the MHV sector. In particular we first compute the $\mathcal{O}(z^{2})$ terms in the OPE in subsection \ref{z2ope}. The invariance of the OPE at this order under the action of $L_{2}$ is proved in subection \ref{ppz2L2}. In subsection \ref{ppznopevir}, we rewrite the terms of $\mathcal{O}(z^{n})$ upto $n=2$ in the positive helicity OPE according to of Virasoro primaries and descendants. 

\subsection{$\mathcal{O}(z^{2})$ terms in positive helicity OPE}
\label{z2ope}

Here we compute the $\mathcal{O}(z^{2})$ terms in the OPE between positive helicity graviton primaries in the MHV sector. The operators which can appear at this order must have dimensions $\left( \frac{7+h}{2}, \frac{\bar{h}-1}{2}\right) $, where $(h,\bar{h}) = \left( \frac{\Delta_{1}+\Delta_{2}+1}{2}, \frac{\Delta_{1}+\Delta_{2}-3}{2}\right)$. We then assume the following ansatz for the $\mathcal{O}(z^{2})$ OPE\footnote{The ansatz \eqref{zsqopedrv1} is not the most general one, since it is possible to write down additional descendants with dimensions $\left( \frac{7+h}{2}, \frac{\bar{h}-1}{2}\right) $. As we will see in this section, the coefficients $x_{1},x_{2},x_{3},x_{4}$ in \eqref{zsqopedrv1}  can be uniquely determined using the underlying symmetry algebras. This justifies why it is sufficient to consider \eqref{zsqopedrv1}.}
\begin{equation}
\label{zsqopedrv1}
\begin{split}
  & G^{+}_{\Delta_{1}} (z,\bar{z}) G^{+}_{\Delta_{2}} (0) \Big|_{\mathcal{O}(z^{2})}  \\
  &=  z^{2} B(\Delta_{1}-1,\Delta_{2}-1) \bigg[ x_{1} \hspace{0.04cm} P_{-4,0} + x_{2}   \hspace{0.04cm}  J^{1}_{-3}P_{-1,-1}  +  x_{3} \hspace{0.04cm} L_{-1}^{2}P_{-2,0} + x_{4} \hspace{0.04cm} L_{-1}P_{-3,0} \bigg] G^{+}_{\Delta_{1}+\Delta_{2}-1} (0) 
\end{split}
\end{equation}

where $x_{1}, x_{2}, x_{3}, x_{4}$ are as of yet undetermined coefficients. Note that using the null state relations
\begin{align}
\label{nstz}
& \left[L_{-1}P_{-2,0} + J^{1}_{-2}P_{-1,-1} -(\Delta_{1}+\Delta_{2}) P_{-3,0}\right] G^{+}_{\Delta_{1}+\Delta_{2}-1} (0) =0  \\ 
\label{nstzb}
  & \left[L_{-1}P_{-1,-1} + 2J^{0}_{-1}P_{-1,-1}-(\Delta_{1}+\Delta_{2})P_{-2,-1} - \bar{L}_{-1}P_{-2,0}\right]G^{+}_{\Delta_{1}+\Delta_{2}-1} (0) =0 
\end{align}

we can express \eqref{zsqopedrv1} purely in terms of $\overline{SL(2,\mathbb{C})}$ current algebra and supertranslation descendants. But for the purpose of determining the coefficients $x_{i}$, it turns out to be more convenient to work with the  form of the OPE given in \eqref{zsqopedrv1}.  

We will now compute the $x_{i}$'s by demanding that both sides of the OPE transform in the same way under the action of the underlying symmetries. For this, let us first act with the $\overline{SL(2,\mathbb{C})}$ current algebra generator  $J^{0}_{1}$ on both sides of \eqref{zsqopedrv1}.

The action of $J^{0}_{1}$ on a conformal primary $\phi_{h,\bar{h}}(z,\bar{z})$ is given by
\begin{equation}
\label{J01act}
\begin{split}
 &  [ J^{0}_{1}, \phi_{h,\bar{h}} (z,\bar{z}) ] = z (\bar{z}\partial_{\bar{z}} + \bar{h}) \phi_{h,\bar{h}} (z,\bar{z}) 
\end{split}
\end{equation}

Using this we get
\begin{equation}
\label{J01act1}
\begin{split}
  &  [ J^{0}_{1}, G^{+}_{\Delta_{1}} (z,\bar{z}) G^{+}_{\Delta_{2}} (0)] = z (\bar{z}\partial_{\bar{z}} + \bar{h}_{1}) G^{+}_{\Delta_{1}} (z,\bar{z}) G^{+}_{\Delta_{2}} (0)
\end{split}
\end{equation}

where $\bar{h}_{1} =(\Delta_{1}-2)/2$.  Then plugging in the $\mathcal{O}(z)$ OPE into the R.H.S. of \eqref{J01act1} yields
\begin{equation}
\label{J01ordzlhs}
\begin{split}
  &  [ J^{0}_{1}, G^{+}_{\Delta_{1}} (z,\bar{z}) G^{+}_{\Delta_{2}} (0)] \Big|_{\mathcal{O}(z^{2})} \\
  & = z^{2}  B(\Delta_{1}-1,\Delta_{2}-1) \hspace{0.04cm} \bar{h}_{1}  \bigg[ \Delta_{1} P_{-3,0} - \frac{(\Delta_{1}-1)}{(\Delta_{1}+\Delta_{2}-2)} J^{1}_{-2}P_{-1,-1} \bigg] G^{+}_{\Delta_{1}+\Delta_{2}-1} (0) 
\end{split}
\end{equation}

Now applying $J^{0}_{1}$ to the R.H.S. of \eqref{zsqopedrv1} and using the commutation relations given in \eqref{comms} we obtain
\begin{equation}
\label{J01ordzrhs}
\begin{split}
&  [ J^{0}_{1}, G^{+}_{\Delta_{1}} (z,\bar{z}) G^{+}_{\Delta_{2}} (0)] \Big|_{\mathcal{O}(z^{2})} \\ 
&  = z^{2} B(\Delta_{1}-1,\Delta_{2}-1) \bigg[ \left(  - \frac{x_{1}}{2} -2 x_{4} + (\Delta_{1}+\Delta_{2}) (\Delta_{1}+\Delta_{2}-4)x_{3} \right)   P_{-3,0} \\
  & \hspace{1.0cm}+ \left(- x_{2} - x_{3}(\Delta_{1}+\Delta_{2}-4) + \frac{x_{4}}{2} \right)   J^{1}_{-2}P_{-1,-1}  \bigg] G^{+}_{\Delta_{1}+\Delta_{2}-1} (0) 
\end{split}
\end{equation}

In arriving at \eqref{J01ordzrhs}, we have also used the null state relation \eqref{nstz}. Now invariance of the OPE under $J^{0}_{1}$ implies that \eqref{J01ordzlhs} and \eqref{J01ordzrhs} should match. As a result we get 
\begin{equation}
\label{zsqrecrl1}
\begin{split}
  &  - \frac{x_{1}}{2} -2 x_{4} + (\Delta_{1}+\Delta_{2}) (\Delta_{1}+\Delta_{2}-4)x_{3} = \bar{h}_{1} \Delta_{1}  
\end{split}
\end{equation}

and
\begin{equation}
\label{zsqrecrl2}
\begin{split}
  &  x_{2} + x_{3}(\Delta_{1}+\Delta_{2}-4)- \frac{x_{4}}{2}  =  \frac{\bar{h}_{1} (\Delta_{1}-1)}{(\Delta_{1}+\Delta_{2}-2)} 
\end{split}
\end{equation}

We need a few more equations to solve for all the $x_{i}$'s. So, let us act with the global spacetime translation generator $P_{0,-1}$ on both sides of \eqref{zsqopedrv1}. $P_{0,-1}$ acts on a outgoing conformal primary $\phi_{h,\bar{h}}(z,\bar{z})$ as follows
\begin{equation}
\label{P0m1act}
\begin{split}
  &  [ P_{0,-1}, \phi_{h,\bar{h}}(z,\bar{z})]  = z \phi_{h+1/2,\bar{h}+1/2}(z,\bar{z})
\end{split}
\end{equation}

\eqref{P0m1act} then implies 
\begin{equation}
\label{P0m1act1}
\begin{split}
  &  [ P_{0,-1}, G^{+}_{\Delta_{1}} (z,\bar{z}) G^{+}_{\Delta_{2}} (0)]  = z G^{+}_{\Delta_{1}+1} (z,\bar{z}) G^{+}_{\Delta_{2}} (0)
\end{split}
\end{equation}

Inserting the $\mathcal{O}(z)$ OPE in the R.H.S. of \eqref{P0m1act1} then gives
\begin{equation}
\label{P0m1ordzlhs}
\begin{split}
  &  [ P_{0,-1}, G^{+}_{\Delta_{1}} (z,\bar{z}) G^{+}_{\Delta_{2}} (0)] \Big|_{\mathcal{O}(z^{2})} \\
  & = z^{2}  B(\Delta_{1},\Delta_{2}-1) \bigg[ (\Delta_{1}+1) P_{-3,0} - \frac{\Delta_{1}}{(\Delta_{1}+\Delta_{2}-1)} J^{1}_{-2}P_{-1,-1} \bigg] G^{+}_{\Delta_{1}+\Delta_{2}} (0) 
\end{split}
\end{equation}

Now evaluating the action of $P_{0,-1}$ on the R.H.S. of \eqref{zsqopedrv1} we get
\begin{equation}
\label{P0m1ordzrhs}
\begin{split}
  &  [ P^{0}_{1}, G^{+}_{\Delta_{1}} (z,\bar{z}) G^{+}_{\Delta_{2}} (0)] \Big|_{\mathcal{O}(z^{2})} \\
&  = z^{2}  B(\Delta_{1}-1,\Delta_{2}-1)  \bigg[ (- x_{2}+x_{4}+ 2x_{3}(\Delta_{1}+\Delta_{2}+1)) P_{-3,0} - 2x_{3} \hspace{0.04cm} J^{1}_{-2}P_{-1,-1}  \bigg] G^{+}_{\Delta_{1}+\Delta_{2}} (0) 
\end{split}
\end{equation}

where we have once again used the null state \eqref{nstz}. Now \eqref{P0m1ordzlhs} and \eqref{P0m1ordzrhs} should match in order for the $\mathcal{O}(z^{2})$ OPE to be invariant under $P_{0,-1}$. Therefore, we get 
\begin{equation}
\label{zsqrecrl3}
\begin{split}
  &  x_{4} + 2x_{3}(\Delta_{1}+\Delta_{2}+1) -x_{2} = \frac{( \Delta_{1}+1 )(\Delta_{1}-1)}{(\Delta_{1}+\Delta_{2}-2)}
\end{split}
\end{equation}

and
\begin{equation}
\label{x3sol}
\begin{split}
  &  x_{3} = \frac{\Delta_{1}(\Delta_{1}-1)}{2 (\Delta_{1}+\Delta_{2}-1)(\Delta_{1}+\Delta_{2}-2)} 
\end{split}
\end{equation}

Equations \eqref{zsqrecrl2} and \eqref{zsqrecrl3} can be simultaneously solved to give
\begin{equation}
\label{x2sol}
\begin{split}
  &  x_{2} = -  \frac{(\Delta_{1}-1)(\Delta_{2}-1)}{ (\Delta_{1}+\Delta_{2}-1)(\Delta_{1}+\Delta_{2}-2)} 
\end{split}
\end{equation}

\begin{equation}
\label{x4sol}
\begin{split}
  &  x_{4} = - \frac{\Delta_{1}(\Delta_{1}-1)}{ (\Delta_{1}+\Delta_{2}-1)(\Delta_{1}+\Delta_{2}-2)} 
\end{split}
\end{equation}

Plugging these results in \eqref{zsqrecrl1} then yields
\begin{equation}
\label{x1sol}
\begin{split}
  &  x_{1} = \frac{\Delta_{1} \Delta_{2} }{ \Delta_{1}+\Delta_{2}-1} 
\end{split}
\end{equation}

The $\mathcal{O}(z^{2})$ OPE \eqref{zsqopedrv1} then takes the form
\begin{equation}
\label{zsqope}
\begin{split}
   G^{+}_{\Delta_{1}} (z,\bar{z}) G^{+}_{\Delta_{2}} (0) \Big|_{\mathcal{O}(z^{2})}  &=  z^{2} B(\Delta_{1}-1,\Delta_{2}-1) \bigg[ x_{1} \hspace{0.04cm} P_{-4,0} + x_{2}   \hspace{0.04cm}  J^{1}_{-3}P_{-1,-1}  \\
   & \hspace{1.5cm}+  x_{3} \left( L_{-1}^{2}P_{-2,0} -2 L_{-1}P_{-3,0} \right) \bigg] G^{+}_{\Delta_{1}+\Delta_{2}-1} (0) 
\end{split}
\end{equation}

where
\begin{equation}
\label{xisols}
\begin{split}
  &  x_{1} = \frac{\Delta_{1} \Delta_{2} }{ \Delta_{1}+\Delta_{2}-1}, \quad  x_{2} = - \frac{(\Delta_{1}-1)(\Delta_{2}-1)}{ (\Delta_{1}+\Delta_{2}-1)(\Delta_{1}+\Delta_{2}-2)} , \\
  & x_{3} = \frac{\Delta_{1}(\Delta_{1}-1)}{2 (\Delta_{1}+\Delta_{2}-1)(\Delta_{1}+\Delta_{2}-2)} 
\end{split}
\end{equation}

\subsection{$L_{2}$ invariance of $\mathcal{O}(z^{2})$ terms in positive helicity OPE }
\label{ppz2L2}

We will now show that the $\mathcal{O}(z^{2})$ OPE computed in the previous section and given by \eqref{zsqope} is invariant under the action of $L_{2}$. In order to do this we start with the Virasoro transformation property of graviton primaries \eqref{L2def} 
\begin{equation}
\label{L2def1}
\begin{split}
  & [ L_{2}, G^{+}_{\Delta_{1}} (z,\bar{z}) G^{+}_{\Delta_{2}} (0)  ] = z^{2} (z\partial_{z} + 3 h_{1}) G^{+}_{\Delta_{1}} (z,\bar{z}) G^{+}_{\Delta_{2}} (0) 
\end{split}
\end{equation}

Substituting the $\mathcal{O}(z^{0}\bar{z}^{0})$ OPE into the R.H.S. of \eqref{L2def1} then yields
\begin{equation}
\label{L2zsqope1}
\begin{split}
   [ L_{2}, G^{+}_{\Delta_{1}} (z,\bar{z}) G^{+}_{\Delta_{2}} (0)  ] \Big|_{\mathcal{O}(z^{2})}  & = z^{2} B(\Delta_{1}-1,\Delta_{2}-1) \ 3h_{1} P_{-2,0} G^{+}_{\Delta_{1}+\Delta_{2}-1} (0)  \\
   & = z^{2} B(\Delta_{1}-1,\Delta_{2}-1) \ \frac{3(\Delta_{1}+2)}{2} P_{-2,0} G^{+}_{\Delta_{1}+\Delta_{2}-1} (0) 
\end{split}
\end{equation}

Now using the commutation relations \eqref{comms} and the values of $x_{1}, x_{2}, x_{3}$ from \eqref{xisols} we find that the action of $L_{2}$ on the R.H.S. of \eqref{zsqope} gives\footnote{In order to derive \eqref{L2zsqope2} we have also used the relation $\left(J^{1}_{-1}P_{-1,-1} - (\Delta-1) P_{-2,0} \right)G^{+}_{\Delta} =0$. }
\begin{equation}
\label{L2zsqope2}
\begin{split}
& z^{2} B(\Delta_{1}-1,\Delta_{2}-1) \left[ \frac{9 x_{1}}{2} + 3x_{2}(\Delta_{1}+\Delta_{2}-2) + 3 x_{3}(\Delta_{1}+\Delta_{2}-2)\right] P_{-2,0} G^{+}_{\Delta_{1}+\Delta_{2}-1} (0) \\
& =   z^{2} B(\Delta_{1}-1,\Delta_{2}-1) \ \frac{3(\Delta_{1}+2)}{2} P_{-2,0} G^{+}_{\Delta_{1}+\Delta_{2}-1} (0) 
\end{split}
\end{equation}

This matches with \eqref{L2zsqope1} and hence exhibits the Virasoro invariance of the $\mathcal{O}(z^{2})$ terms in the OPE \eqref{zsqope}. Let us again note here that as in the $\mathcal{O}(z\bar{z})$ OPE, the $L_{-2}$ descendant of $G^{+}_{\Delta_{1}+\Delta_{2}}$ does not appear in the OPE \eqref{zsqope}. Therefore our calculation here is insensitive to the central charge of the Virasoro algebra and Virasoro invariance of \eqref{zsqope} continues to hold for any value of the central charge. 


\subsection{Reorganising OPE in terms of Virasoro primaries and descendants}
\label{ppznopevir}

In section \ref{virrep1} we showed that the $\mathcal{O}(\bar{z}z^{n})$ terms in the positive helicity graviton OPE in the MHV sector can be reorganised according to Virasoro primaries and descendants. Here we provide another example of this reorganisation by considering terms of $\mathcal{O}(z^{n})$ in the OPE. So, let us  begin with the following set of terms in the OPE,
\begin{equation}
\label{ordznppope}
\begin{split}
  & G^{+}_{\Delta_{1}} (z,\bar{z}) G^{+}_{\Delta_{2}} (0) \Big|_{MHV} \\
  & =  B(\Delta_{1}-1,\Delta_{2}-1) \bigg[  P_{-2,0} + z \big( d_{1} P_{-3,0} + d_{2} J^{1}_{-2}P_{-1,-1} \big)  \\
 & + z^{2} \big(x_{1} P_{-4,0} + x_{2} J^{1}_{-3}P_{-1,-1} +  x_{3} \left( L_{-1}^{2}P_{-2,0} - 2 L_{-1}P_{-3,0} \right) \big) + \mathcal{O}\left(z^{3}\right) \bigg] G^{+}_{\Delta_{1}+\Delta_{2}-1} (0) + \cdots
\end{split}
\end{equation}

where $d_{1},d_{2}$ in \eqref{ppopedi} were obtained using symmetries in \cite{Banerjee:2020zlg} and are given by
\begin{equation}
\label{ppopedi}
\begin{split}
  & d_{1} = \Delta_{1}, \quad d_{2} = - \frac{\Delta_{1}-1}{\Delta_{1}+\Delta_{2}-2}
\end{split}
\end{equation}

$x_{1},x_{2}, x_{3}$ are given in \eqref{xisols}. Now \eqref{ordznppope} can also be written as 
\begin{equation}
\label{ppvirrep1}
\begin{split}
  & G^{+}_{\Delta_{1}} (z,\bar{z}) G^{+}_{\Delta_{2}} (0) \Big|_{MHV} \\
    & = B(\Delta_{1}-1,\Delta_{2}-1) \bigg[ C_{12\psi_{1}} \left( 1 + z \hspace{0.04cm} \beta_{\psi_{1}}^{(1)} L_{-1}+ z^{2}\beta_{1}^{(1,1)} L_{-1}^{2}+ z^{2}\beta_{\psi_{1}}^{(2)}L_{-2} \right) \psi_{1}(0) \\
   & + z \hspace{0.03cm} C_{12\psi_{2}} \left( 1 + z \hspace{0.04cm}  \beta_{\psi_{2}}^{(1)} L_{-1} \right)  \psi_{2}(0) + z^{2} \hspace{0.03cm} C_{12\psi_{3}} \psi_{3}(0) +  \mathcal{O}\left(z^{3}\right)\bigg] + \cdots
\end{split}
\end{equation}

where $\psi_{i}(0)$ with $i=1,2,3$ are Virasoro primaries for arbitrary values of the central charge. Their explicit expressions are
\begin{equation}
\label{psi}
\begin{split}
   &   \psi_{1}(0) = P_{-2,0}G^{+}_{\Delta_{1}+\Delta_{2}-1} (0), \quad  \psi_{2}(0) = \left(P_{-3,0}- \frac{3}{(\Delta_{1}+\Delta_{2}+4)} L_{-1}P_{-2,0}\right)G^{+}_{\Delta_{1}+\Delta_{2}-1} (0), \\
  &  \psi_{3}(0) =  \bigg(  P_{-4,0} + b_{1} J^{1}_{-3}P_{-1,-1} + b_{2} L_{-1}P_{-3,0} + b_{3}L_{-1}^{2}P_{-2,0}  + b_{4} L_{-2}P_{-2,0}\bigg)G^{+}_{\Delta_{1}+\Delta_{2}-1} (0) 
    \end{split}
\end{equation}

where the $b_{i}$'s in \eqref{psi} are given by
\begin{equation}
\label{bis}
\begin{split}
  & b_{1} = - \frac{(\Delta_{1}-1)(\Delta_{2}-1)}{C_{12\psi_{3}}  (\Delta_{1}+\Delta_{2}-1) (\Delta_{1}+\Delta_{2}-2)} \\
 & b_{2} =  \frac{1}{C_{12\psi_{3}} } \bigg(  -\frac{(\Delta_{1}-1)(\Delta_{2}-1)}{(\Delta_{1}+\Delta_{2}-1)(\Delta_{1}+\Delta_{2}-2)} - C_{12\psi_{2}} \  \beta_{\psi_{2}}^{(1)} \bigg) \\
 & b_{3} =  \frac{1}{C_{12\psi_{3}} } \bigg( \frac{(\Delta_{1}-1)(\Delta_{2}-1)}{2(\Delta_{1}+\Delta_{2}-1)(\Delta_{1}+\Delta_{2}-2)} - C_{12\psi_{1}} \ \beta_{\psi_{1}}^{(1,1)} + \frac{3 \hspace{0.03cm} C_{12\psi_{2}} \ \beta_{\psi_{2}}^{(1)}}{(\Delta_{1}+\Delta_{2}+4)}  \bigg) \\
 & b_{4} = - \frac{C_{12\psi_{1}} \ \beta_{\psi_{1}}^{(2)}}{C_{12\psi_{3}}}
  \end{split}
\end{equation}

The coefficients $C_{12\psi_{i}}$ in \eqref{ppvirrep1} are given by
 \begin{equation}
\label{C12p} 
\begin{split}
& C_{12\psi_{1}} = 1, \quad C_{12\psi_{2}} = \frac{\Delta_{2}-\Delta_{1}}{\Delta_{1}+\Delta_{2}-2}   , \quad   C_{12\psi_{3}}  = \frac{\Delta_{1}\Delta_{2}}{\Delta_{1}+\Delta_{2}-1}
\end{split}
\end{equation}

Note that the  $C_{12\psi_{i}}$'s cannot be determined by using only the Virasoro algebra. The coefficients $\beta^{(\vec{k})}_{\psi_{i}}$ in \eqref{ppvirrep1} are uniquely fixed by the Virasoro algebra. The explicit formulae for the $\beta^{(\vec{k})}_{\psi_{i}}$'s take the same form as in \eqref{betas} with $h_{\phi_{i}}$ replaced by $h_{\psi_{i}}$, where $h_{\psi_{i}}$ denotes the holomorphic weight of the Virasoro primary $\psi_{i}$ appearing in the R.H.S. of \eqref{ppvirrep1}.

\section{Virasoro invariance of mixed helicity OPE}

In this section of the Appendix, we consider the celestial OPE between an outgoing positive helicity graviton primary and an outgoing negative helicity graviton primary in the MHV sector. Here we show that this mixed helicity OPE is also Virasoro invariant. In order to furnish a  non-trivial check of Virasoro invariance we first compute the $\mathcal{O}(z\bar{z})$ terms in the mixed helicity OPE in subsection \ref{pmopezzb}. This is followed by an explicit check of the invariance of this OPE under the action of $L_{2}$ in subsection \ref{pmopeL2}. In subsection \ref{pmopevir} it is shown that the mixed helicity OPE can be organised according to Virasoro representations.  

\subsection{$\mathcal{O}(z\bar{z}) $ term in mixed helicity OPE using symmetries}
\label{pmopezzb}

The operators which can appear at $\mathcal{O}(z\bar{z}) $ in the mixed helicity OPE should have weights $\left( \frac{3+h}{2}, \frac{\bar{h}+1}{2}\right) $, where $(h,\bar{h}) = \left( \frac{\Delta_{1}+\Delta_{2}-3}{2}, \frac{\Delta_{1}+\Delta_{2}+1}{2}\right)$. Let us then consider the OPE at this order to take the following form
\begin{equation}
\label{pmopezzbar}
\begin{split}
  & G^{+}_{\Delta_{1}} (z,\bar{z}) G^{-}_{\Delta_{2}} (0) \Big|_{\mathcal{O}(z\bar{z})}  \\
  & = z\bar{z}  \ B(\Delta_{1}-1,\Delta_{2}+3) \bigg[ \bigg(\alpha_{1}J^{0}_{-2}P_{-1,-1} + \alpha_{2}P_{-3,-1} +\alpha_{3} J^{0}_{-1}P_{-2,-1} + \alpha_{4} (J^{0}_{-1})^{2}P_{-1,-1} \\
  & + \alpha_{5} \bar{L}_{-1}P_{-3,0} + \alpha_{6} \bar{L}_{-1} J^{0}_{-1}P_{-2,0} + \alpha_{7}J^{-1}_{-1}P_{-2,0} +\alpha_{8} \bar{L}_{-1} J^{1}_{-2}P_{-1,-1}\bigg) G^{-}_{\Delta_{1}+\Delta_{2}-1} (0)  \\
  & + \bigg(\alpha_{9} P_{-2,-1}^{2}+ \alpha_{10} \bar{L}_{-1}P_{-2,-1}P_{-2,0} + \alpha_{11}\bar{L}_{-1}^{2}P_{-2,0}^{2}\bigg) G^{-}_{\Delta_{1}+\Delta_{2}-2} (0) \bigg]
\end{split}
\end{equation}

where $\alpha_{i}, i=1,2, \ldots, 11$ are as of yet unknown coefficients. In order to determine these using symmetries we follow the same procedure as in Section \ref{z2ope}. We impose that both sides of the OPE  \eqref{pmopezzbar} transform identically under the action of supertranslations and the $\overline{SL(2,\mathbb{C})}$ current algebra. This yields a system of equations for the $\alpha_{i}$'s which can then be simultaneously solved. In particular we will apply below the generators $P_{0,-1}, P_{-1,0}$ and $J^{1}_{0}$ for deriving these equations. 

First let us consider acting with $P_{0,-1}$. Demanding both sides of \eqref{pmopezzbar} to transform in the same fashion under the action of $P_{0,-1}$ leads to 
\begin{align}
\label{recrel1}
&\alpha_{4} = - \frac{2 \Delta_{1}(\Delta_{1}-1)}{(\Delta_{1}+\Delta_{2}+2)(\Delta_{1}+\Delta_{2}+3)} \\
& \nonumber \\
\label{recrel2}
&-\frac{\alpha_{1}}{2}- \frac{\alpha_{3}}{2} +  \frac{\alpha_{4}}{4} = - \frac{(\Delta_{1}+1)(\Delta_{1}-1)}{\Delta_{1}+\Delta_{2}+2} \\
& \nonumber \\
\label{recrel3}
& \alpha_{6}+ 2\alpha_{8} =0
\end{align}

Next we act with $P_{-1,0}$ on both sides of the OPE \eqref{pmopezzbar}.  In this case we obtain
\begin{align}
\label{recrel4}
& \alpha_{6} = - \alpha_{4} = \frac{2 \Delta_{1}(\Delta_{1}-1)}{(\Delta_{1}+\Delta_{2}+2)(\Delta_{1}+\Delta_{2}+3)} \\
& \nonumber \\
\label{recrel5}
 & \alpha_{8} = - \frac{ \Delta_{1}(\Delta_{1}-1)}{(\Delta_{1}+\Delta_{2}+2)(\Delta_{1}+\Delta_{2}+3)} \\
 & \nonumber \\
\label{recrel6}
& \alpha_{11} = - \frac{\alpha_{6}}{4} = - \frac{ \Delta_{1}(\Delta_{1}-1)}{2(\Delta_{1}+\Delta_{2}+2)(\Delta_{1}+\Delta_{2}+3)} \\
 & \nonumber \\
\label{recrel7}
& \frac{\alpha_{1}}{2} + \frac{\alpha_{4}}{4} +  \alpha_{5} -\alpha_{8}= \frac{(\Delta_{1}+1)(\Delta_{1}-1)}{\Delta_{1}+\Delta_{2}+2}  \\ 
& \nonumber \\
\label{recrel8}
& \frac{\alpha_{3}}{2}  -   \frac{\alpha_{6}}{2} + \alpha_{7} + \alpha_{10}  = 0 
\end{align}

Finally invariance of the OPE under the action of  $J^{1}_{0}$ yields
\begin{align}
\label{recrel9}
& \alpha_{1} + \alpha_{4}+ (2\bar{h}-1) \alpha_{8} =  - \frac{2\bar{h}_{1}(\Delta_{1}-1)}{\Delta_{1}+\Delta_{2}+2} \\
& \nonumber \\
\label{recrel10}
& \alpha_{2} + \alpha_{3}+ (2\bar{h}-1) \alpha_{5} =  2\bar{h}_{1} \Delta_{1} \\
& \nonumber \\
\label{recrel11}
& \alpha_{3} +  2(\Delta+3)  \alpha_{4}+ (2\bar{h}-1)  \alpha_{6} +2  \alpha_{7} = 0\\
& \nonumber \\
\label{recrel12}
& (\Delta+2) \alpha_{3} +  2 \alpha_{9}+ (2\bar{h}-1) \alpha_{10} = 0\\
& \nonumber \\
\label{recrel13}
& (\Delta+2) \alpha_{6} +   \alpha_{10}+ 4(\bar{h}-1) \alpha_{11} = 0
\end{align}

The above system of equations can be solved to uniquely determine all the $\alpha_{i}$'s. These are as follows
\begin{equation}
\label{alphais}
\begin{split}
& \alpha_{1} = \frac{(\Delta_{1}-1) (\Delta_{1}+2 \Delta_{2}+6)}{(\Delta_{1}+\Delta_{2}+2) (\Delta_{1}+\Delta_{2}+3)}, \quad \alpha_{2} = -\Delta_{1}, \quad \alpha_{3} = \frac{2\Delta_{1}(\Delta_{1}-1)}{\Delta_{1}+\Delta_{2}+2}, \\ 
&  \alpha_{4} = - \frac{2 \Delta_{1}(\Delta_{1}-1)}{(\Delta_{1}+\Delta_{2}+2)(\Delta_{1}+\Delta_{2}+3)}, \quad \alpha_{5} =  \frac{ \Delta_{1}(\Delta_{1}-1)}{\Delta_{1}+\Delta_{2}+2}, \quad \alpha_{6} =  \frac{2 \Delta_{1}(\Delta_{1}-1)}{(\Delta_{1}+\Delta_{2}+2)(\Delta_{1}+\Delta_{2}+3)} \\ 
& \alpha_{7} =  \frac{ \Delta_{1}(\Delta_{1}-1)}{(\Delta_{1}+\Delta_{2}+2)(\Delta_{1}+\Delta_{2}+3)}, \quad \alpha_{8} = - \frac{ \Delta_{1}(\Delta_{1}-1)}{(\Delta_{1}+\Delta_{2}+2)(\Delta_{1}+\Delta_{2}+3)} \\ 
& \alpha_{9} = -\frac{\Delta_{1}(\Delta_{1}-1)}{2}, \quad \alpha_{10} = - \frac{\Delta_{1}(\Delta_{1}-1)}{\Delta_{1}+\Delta_{2}+2}, \quad \alpha_{11} = - \frac{ \Delta_{1}(\Delta_{1}-1)}{2(\Delta_{1}+\Delta_{2}+2)(\Delta_{1}+\Delta_{2}+3)} 
\end{split}
\end{equation}




\subsection{$L_{2}$ invariance of $\mathcal{O}(z\bar{z})$ terms in mixed helicity OPE}
\label{pmopeL2}

We shall now verify that the mixed helicity OPE at $\mathcal{O}(z\bar{z})$ \eqref{pmopezzbar} is invariant under $L_{2}$.  For this we act with $L_{2}$ on both sides of  \eqref{pmopezzbar}. From the L.H.S. we get
\begin{equation}
\label{pmL2zzbar1}
\begin{split}
   [L_{2}, G^{+}_{\Delta_{1}} (z,\bar{z}) G^{-}_{\Delta_{2}} (0)]\Big|_{\mathcal{O}(z\bar{z})}  & = - z\bar{z}  \ B(\Delta_{1}-1,\Delta_{2}+3) \ \frac{(3\Delta_{1}+4)}{2}P_{-1,-1}G^{-}_{\Delta_{1}+\Delta_{2}-1} (0)
\end{split}
\end{equation}

The action of $L_{2}$ on the R.H.S. of \eqref{pmopezzbar} evaluates to 
\begin{equation}
\label{pmL2zzbar2}
\begin{split}
 & z\bar{z}  \ B(\Delta_{1}-1,\Delta_{2}+3) \left[ (\Delta_{1}+\Delta_{2}+2) \alpha_{1} + \frac{7 \alpha_{2}}{2} + \alpha_{5} - \frac{\alpha_{6}}{2}\right]P_{-1,-1}G^{-}_{\Delta_{1}+\Delta_{2}-1} (0)  
\end{split}
\end{equation}

Using the values of $\alpha_{1}, \alpha_{2}, \alpha_{5}, \alpha_{6}$ from \eqref{alphais}, we find that \eqref{pmL2zzbar2} simplifies to
\begin{equation}
\label{pmL2zzbar3}
\begin{split}
  &  - z\bar{z}  \ B(\Delta_{1}-1,\Delta_{2}+3) \ \frac{(3\Delta_{1}+4)}{2}P_{-1,-1}G^{-}_{\Delta_{1}+\Delta_{2}-1} (0)
\end{split}
\end{equation}

This matches with \eqref{pmL2zzbar1}  and thereby proves that the $\mathcal{O}(z\bar{z})$ term in the mixed helicity OPE is Virasoro invariant. As in the case of the positive helicity OPE, the central charge of the Virasoro algebra does not play any role here.  


\subsection{Reorganising OPE in terms of Virasoro primaries and descendants}
\label{pmopevir}

In the MHV sector, the mixed helicity graviton OPE takes the following form
\begin{equation}
\label{pmope}
\begin{split}
  &  G^{+}_{\Delta_{1}} (z,\bar{z}) G^{-}_{\Delta_{2}} (0) \Big|_{MHV}  \\
    & = \frac{\bar{z}}{z} \   B(\Delta_{1}-1,\Delta_{2}+3) \bigg[ - P_{-1,-1} G^{-}_{\Delta_{1}+\Delta_{2}-1} (0)   + z \big(\rho_{1} \hspace{0.04cm} J^{0}_{-1}P_{-1,-1} + \rho_{2} P_{-2,-1}  \big) G^{-}_{\Delta_{1}+\Delta_{2}-1} (0) \\
 & \hspace{1.0cm} + z^{2} \bigg( \alpha_{1} J^{0}_{-2}P_{-1,-1} + \alpha_{2} P_{-3,-1}  + \alpha_{5} \left( 2J^{0}_{-1}P_{-2,-1}   + \bar{L}_{-1}P_{-3,0} \right)\\ 
 & \hspace{1.0cm}+ \frac{\alpha_{6}}{2} \big( -2 (J^{0}_{-1})^{2}P_{-1,-1}  + 2 \bar{L}_{-1}J^{0}_{-1}P_{-2,0} + J^{-1}_{-1}P_{-2,0} - \bar{L}_{-1}J^{1}_{-2}P_{-1,-1} \big)\bigg) G^{-}_{\Delta_{1}+\Delta_{2}-1} (0) \\ 
  & \hspace{1.0cm}  + z^{2} \big( \alpha_{9} P_{-2,-1}^{2}  + \alpha_{10} \bar{L}_{-1}P_{-2,-1}P_{-2,0}  + \alpha_{11} \bar{L}_{-1}^{2}P_{-2,0}^{2}\big) G^{-}_{\Delta_{1}+\Delta_{2}-2}(0) + \mathcal{O}\left(z^{3}\right) \bigg] \\ 
 & + B(\Delta_{1}-1,\Delta_{2}+3) \bigg[  P_{-2,0}G^{-}_{\Delta} (0)  + z \bigg( \eta_{1} P_{-3,0} + \eta_{2} J^{1}_{-2}P_{-1,-1} \bigg) G^{-}_{\Delta_{1}+\Delta_{2}-1} (0) +\mathcal{O}\left(z^{2}\right) \bigg] + \cdots
\end{split}
\end{equation}

 where  $\rho_{1}, \rho_{2}$ and $\eta_{1}, \eta_{2}$ are given by
\begin{equation}
\label{rhoietai}
\begin{split}
 & \rho_{1} = \frac{2(\Delta_{1}-1)}{\Delta_{1}+\Delta_{2}+2}, \quad \rho_{2} =  - \Delta_{1}, \quad \eta_{1} =  \Delta_{1} , \quad \eta_{2} =  - \frac{\Delta_{1}-1}{\Delta_{1}+\Delta_{2}+2}
\end{split}
\end{equation}

The values of $\rho_{1},\rho_{2}$ were computed  previously in \cite{Banerjee:2020zlg}. Although we have not shown in this paper, $\eta_{1}, \eta_{2}$ can also be easily obtained from symmetries. The $\alpha_{i}$'s are given by \eqref{alphais}. The dots in \eqref{pmope} denote further terms which we have suppressed for simplicity. 

Now it is again possible to rearrange \eqref{pmope} in terms of Virasoro primaries and descendants, so as to make the Virasoro invariance of the OPE manifest.  Carrying out this reorganisation we find the following result
\begin{equation}
\label{pmvirrep}
\begin{split}
  & G^{+}_{\Delta_{1}} (z,\bar{z}) G^{-}_{\Delta_{2}} (0) \Big|_{MHV} \\
  & = \frac{\bar{z}}{z} \  B(\Delta_{1}-1,\Delta_{2}+3) \bigg[ C_{12\varphi_{1}}  \left( 1 + z \hspace{0.04cm} \beta_{\varphi_{1}}^{(1)} L_{-1}+ z^{2}\beta_{\varphi_{1}}^{(1,1)} L_{-1}^{2}+ z^{2}\beta_{\varphi_{1}}^{(2)}L_{-2} \right) \varphi_{1}(0)  \\
   &  +  z \hspace{0.03cm}  C_{12\varphi_{2}}  \left( 1 + z \hspace{0.04cm}  \beta_{\varphi_{2}}^{(1)} L_{-1} \right) \varphi_{2}(0)   + z \hspace{0.03cm} C_{12\varphi_{3}}  \left( 1 + z \hspace{0.04cm}  \beta_{\varphi_{3}}^{(1)} L_{-1} \right) \varphi_{3}(0)  + z^{2} \hspace{0.03cm} C_{12\varphi_{4}}   \varphi_{4}(0) + \mathcal{O}\left(z^{3}\right) \bigg] \\
   & + B(\Delta_{1}-1,\Delta_{2}+3) \bigg[ C_{12\Omega_{1}} \left( 1 + z \hspace{0.04cm} \beta_{\Omega_{1}}^{(1)} L_{-1}\right) \Omega_{1}(0)  + z \hspace{0.03cm} C_{12\Omega_{2}}  \hspace{0.04cm} \Omega_{2}(0)  + \mathcal{O}\left(z^{2}\right) \bigg] + \cdots
\end{split}
\end{equation}

where the operators denoted by $\varphi_{i}(0)$ and $\Omega_{i}(0)$ are Virasoro primaries (for any value of the central charge). Their explicit expressions are given by \eqref{pmvirprims1} and \eqref{pmvirprims2}. The coefficients $C_{12\varphi_{i}}$ and $C_{12\Omega{i}}$ are 
\begin{equation}
\label{pmprcoeffs}
\begin{split}
  & C_{12\varphi_{1}}  = - 1, \quad C_{12\varphi_{2}} = - \Delta_{1},  \quad C_{12\varphi_{3}} = - \Delta_{1}, \quad C_{12\Omega_{1}} = 1, \quad C_{12\Omega_{2}} = \Delta_{1}
\end{split}
\end{equation}

Once again it is worth noting that Virasoro symmetry alone does not fix the values given in \eqref{pmprcoeffs}. The coefficients $\beta^{(\vec{k})}_{\varphi_{i}}$'s and $\beta^{(\vec{k})}_{\Omega_{i}}$'s in \eqref{pmvirrep}, which are uniquely determined by the Virasoro algebra, are given by similar expressions as in \eqref{betas}. Let us finally note  the expressions for the Virasoro primaries $\varphi_{i}(0), \Omega_{i}(0)$. 
\begin{equation}
\label{pmvirprims1}
\begin{split}
  &   \varphi_{1}(0) = P_{-1,-1}G^{-}_{\Delta_{1}+\Delta_{2}-1} (0) , \\
  & \varphi_{2}(0) = \left(P_{-2,-1}- \frac{2(\Delta_{1}-1)}{\Delta_{1}(\Delta_{1}+\Delta_{2}+2)} J^{0}_{-1}P_{-1,-1}-\frac{(\Delta_{1}+1)}{\Delta_{1}(\Delta_{1}+\Delta_{2}-2)} L_{-1}P_{-1,-1}\right)G^{-}_{\Delta_{1}+\Delta_{2}-1} (0)\\
  & \varphi_{3}(0) = \bigg[ P_{-3,-1} - \frac{(\Delta_{1}-1)(\Delta_{1}+2\Delta_{2}+6)}{\Delta_{1}(\Delta_{1}+\Delta_{2}+3)} J^{0}_{-2}P_{-1,-1}  \\
  & \hspace{1.5cm}  - \frac{ (\Delta_{1}-1)}{\Delta_{1}+\Delta_{2}+2} \left( 2J^{0}_{-1}P_{-2,-1} + \bar{L}_{-1}P_{-3,0}  \right) \\
 & \hspace{1.5cm} + \frac{ (\Delta_{1}-1)}{(\Delta_{1}+\Delta_{2}+2)(\Delta_{1}+\Delta_{2}+3)}  \bigg( 2 (J^{0}_{-1})^{2}P_{-1,-1} - 2 \bar{L}_{-1}J^{0}_{-1}P_{-2,0} \\
 &  \hspace{7.5cm}-  J^{-1}_{-1}P_{-2,0} + \bar{L}_{-1}J^{1}_{-2}P_{-1,-1}  \bigg) \bigg] G^{-}_{\Delta_{1}+\Delta_{2}-1}(0) \\ 
 & \hspace{1.5cm}  + \bigg(  \frac{(\Delta_{1}-1)}{2} P_{-2,-1}^{2}  + \frac{(\Delta_{1}-1)}{\Delta_{1}+\Delta_{2}+2} \ \bar{L}_{-1}P_{-2,-1}P_{-2,0}  \\
 &  \hspace{2.5cm}  + \frac{(\Delta_{1}-1)}{2(\Delta_{1}+\Delta_{2}+2)(\Delta_{1}+\Delta_{2}+3)} \bar{L}_{-1}^{2}P_{-2,0}^{2}\bigg) G^{-}_{\Delta_{1}+\Delta_{2}-2}(0) \\ 
  & \hspace{1.0cm}  - \frac{(\Delta_{1}-4)(4\Delta_{1}+3)}{2\Delta_{1}(\Delta_{1}+\Delta_{2}-2)(4\Delta_{1}+\Delta_{2}-13)} L_{-1}^{2}P_{-1,-1}G^{-}_{\Delta_{1}+\Delta_{2}-1}(0)  \\
  &  \hspace{1.0cm}  - \frac{(\Delta_{1}(3\Delta_{2}-8)+4\Delta_{2}-10)}{\Delta_{1}(\Delta_{1}+\Delta_{2}-2)(4\Delta_{1}+\Delta_{2}-13)} L_{-2}P_{-1,-1}G^{-}_{\Delta_{1}+\Delta_{2}-1}(0) - \frac{(\Delta_{1}+2)}{\Delta_{1}+\Delta_{2}} L_{-1}\varphi_{2}(0)  
    \end{split}
\end{equation}

and
\begin{equation}
\label{pmvirprims2}
\begin{split}
  &  \Omega_{1}(0) = P_{-2,0}G^{-}_{\Delta_{1}+\Delta_{2}-1} (0), \\
  &  \Omega_{2}(0) = \left(P_{-3,0}- \frac{(\Delta_{1}+2)}{\Delta_{1}(\Delta_{1}+\Delta_{2})} L_{-1}P_{-2,0}-\frac{(\Delta_{1}-1)}{\Delta_{1}(\Delta_{1}+\Delta_{2}+2)} J^{1}_{-2}P_{-1,-1}\right)G^{-}_{\Delta_{1}+\Delta_{2}-1} (0) 
    \end{split}
\end{equation}

\end{document}